\documentstyle[epsfig]{aipproc}
\begin{document}
\title{Superluminal Recession Velocities}

\author{Tamara M. Davis and Charles H. Lineweaver}
\address{University of New South Wales
\\Sydney, Australia 2052}

\maketitle
\vspace{-1cm}
\begin{abstract}
Hubble's Law, $v=HD$ (recession velocity is proportional to distance), is a theoretical result derived from the 
Friedmann-Robertson-Walker metric. $v=HD$ applies at least as far as the particle horizon and in principle for all distances.
Thus, galaxies with distances greater than $D=c/H$ are receding from us with velocities greater than the speed of light and
superluminal recession is a fundamental part of the general relativistic description of the expanding universe.
This apparent contradiction of special relativity (SR) is often  mistakenly remedied by converting redshift to velocity
using SR. 
Here we show that galaxies with recession velocities faster than the speed of light are observable and that in 
all viable cosmological models, galaxies above a redshift of three are receding superluminally. 

\end{abstract}
\vspace{-10pt}
\section*{Two Kinds of Velocity}
%\vspace{-10pt}

Despite the efforts of Murdoch~\cite{C:tdavis:murdoch77}, Harrison~\cite{C:tdavis:harrison81,C:tdavis:harrison91,C:tdavis:harrison93}, Stuckey~\cite{C:tdavis:stuckey92}, Ellis \& Rothman~\cite{C:tdavis:ellis93} and Kiang~\cite{C:tdavis:kiang97} 
there is much confusion about superluminal recession in the literature~\cite{C:tdavis:davis00}.
Special relativistic calculations are often incorrectly invoked in a cosmological context and recession velocities with an upper limit of c are obtained.
%For example, special relativistic corrections are often invoked when recession velocities of the order $c$ are obtained.  
A common, but mistaken line of reasoning goes as follows: 
{\bf 1.} The most distant object we can observe has an infinite redshift. 
{\bf 2.} An infinite redshift implies a velocity equal to the speed of light. 
{\bf 3.} Therefore the most distant object we can see has a velocity of $c$.  
This logic falters because the second premise is invalid: 
({\bf 2.}) would only be true if the cosmological redshift were a special relativistic rather than a general relativistic effect.  
Since redshifts and recession velocities are central to the idea of an expanding universe, the distinction between recession 
and peculiar velocity is fundamental.
We define the velocities and redshifts mentioned above and present spacetime diagrams showing how light can reach us from 
regions of space which are receding superluminally.  We have generalised the solutions to arbitrary $(\Omega_M,\Omega_{\Lambda})$.

\begin{figure}[!b] % fig 1
\centerline{\epsfig{file=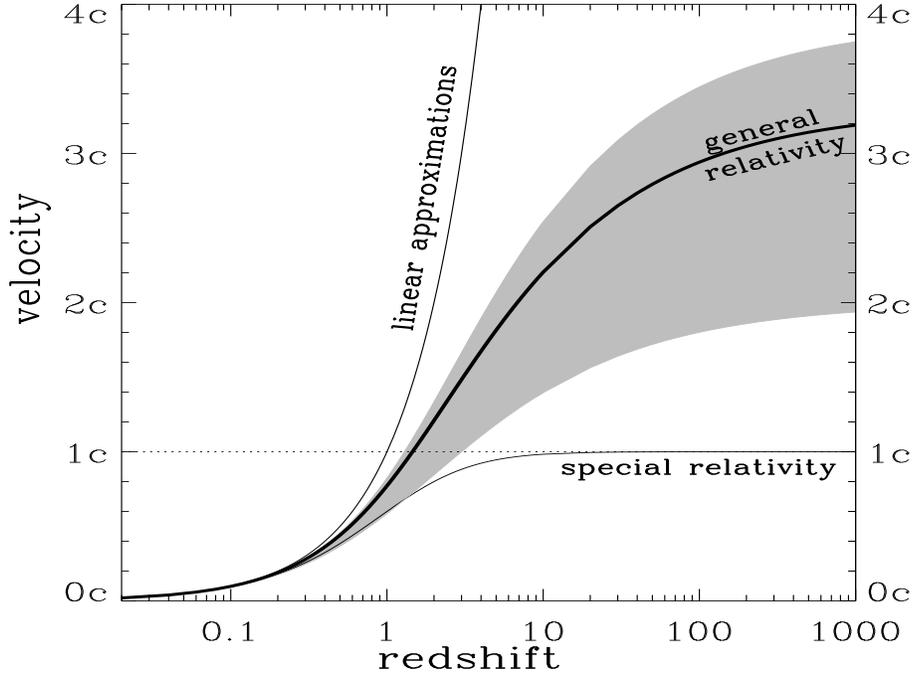,height=3.9in,width=4.8in}}
\vspace{-20pt}
\caption{Velocity is plotted as a function of redshift for both special relativity (Eq.~\ref{E:tdavis:SR}) and the general relativistic (GR) Friedmann-Robertson-Walker cosmology (Eq.~\ref{E:tdavis:GR}). 
The SR conversion has an upper limit of $c$.  The GR conversion is model dependent (hence the grey band), and time dependent (present day recession velocities are shown).  The thick line is an $(\Omega_M,\Omega_{\Lambda})=(0.3,0.7)$ model for which recession velocity exceeds the speed of light for all redshifts $z>1.46$.  
The upper and lower limits of the grey band are $(\Omega_M,\Omega_{\Lambda})=(0.2,0.8)$ and $(1,0)$ respectively.  In all cosmological models within this range, objects above a redshift of three are receding superluminally. 
Unfortunately adding to the confusion,  both expressions share the same low redshift linear approximation: $v \approx c\;z$.  This is why Hubble could measure a linear distance-redshift relation. These redshift-velocity conversions do not depend on $H$. Notice that at the particle horizon ($z \rightarrow \infty$) the velocity of recession is about 3 times the speed of light.}
\label{F:tdavis:fig1}
\end{figure}
%\gtrsim

The Friedmann-Robertson-Walker (FRW) metric~\cite{C:tdavis:peacock99} describes an homogeneous, isotropic universe.
We are interested in only the radial components:
$ds^2 =-c^2dt^2+a(t)^2 d\chi^2$.
Radial distance along a constant time slice ($dt=0$) is given by:
$D=a\chi$.
Differentiating this physical or proper distance gives a two component velocity:\vspace{-2mm}
\begin{eqnarray}
v_{tot}&=& \dot{a}\chi + a\dot{\chi} \nonumber\\
 &=& v_{rec} + v_{pec}
\end{eqnarray}
Recession velocity ($v_{rec}$) is the velocity in Hubble's law since Hubble's constant $H=\dot{a}/a$.  
Peculiar velocity ($v_{pec}$) is unassociated with the expansion of the universe and 
corresponds more closely to common usage of the word `velocity'.
For example setting $ds=0$ for a photon in the FRW metric gives a radial peculiar velocity $a\dot{\chi}_{\gamma}=c$. 
The conversion from cosmological redshift to recession velocity is different from the conversion from an SR Doppler shift 
to peculiar velocity (Fig.~\ref{F:tdavis:fig1}):
\begin{eqnarray}                             
v_{pec}(z) =& c \: \frac{(1+ z)^2-1}{(1 + z)^2+1}&\qquad \mbox{\hspace{2mm} SR Doppler shift} \label{E:tdavis:SR} \\ 
v_{rec}(z) =&  \dot{a}(z)\;\int_o^z\frac{c\:dz^{\prime}}{H(z^{\prime})}&\qquad \mbox{GR cosmological redshift}\label{E:tdavis:GR}
\end{eqnarray}
Equation (\ref{E:tdavis:GR}) converts high redshifts into superluminal recession velocities but
does not contradict special relativity because nothing ever overtakes a photon and all observers 
measure local photons to be traveling at $c$.

\begin{figure}[!h] % fig 2
\centerline{\epsfig{file=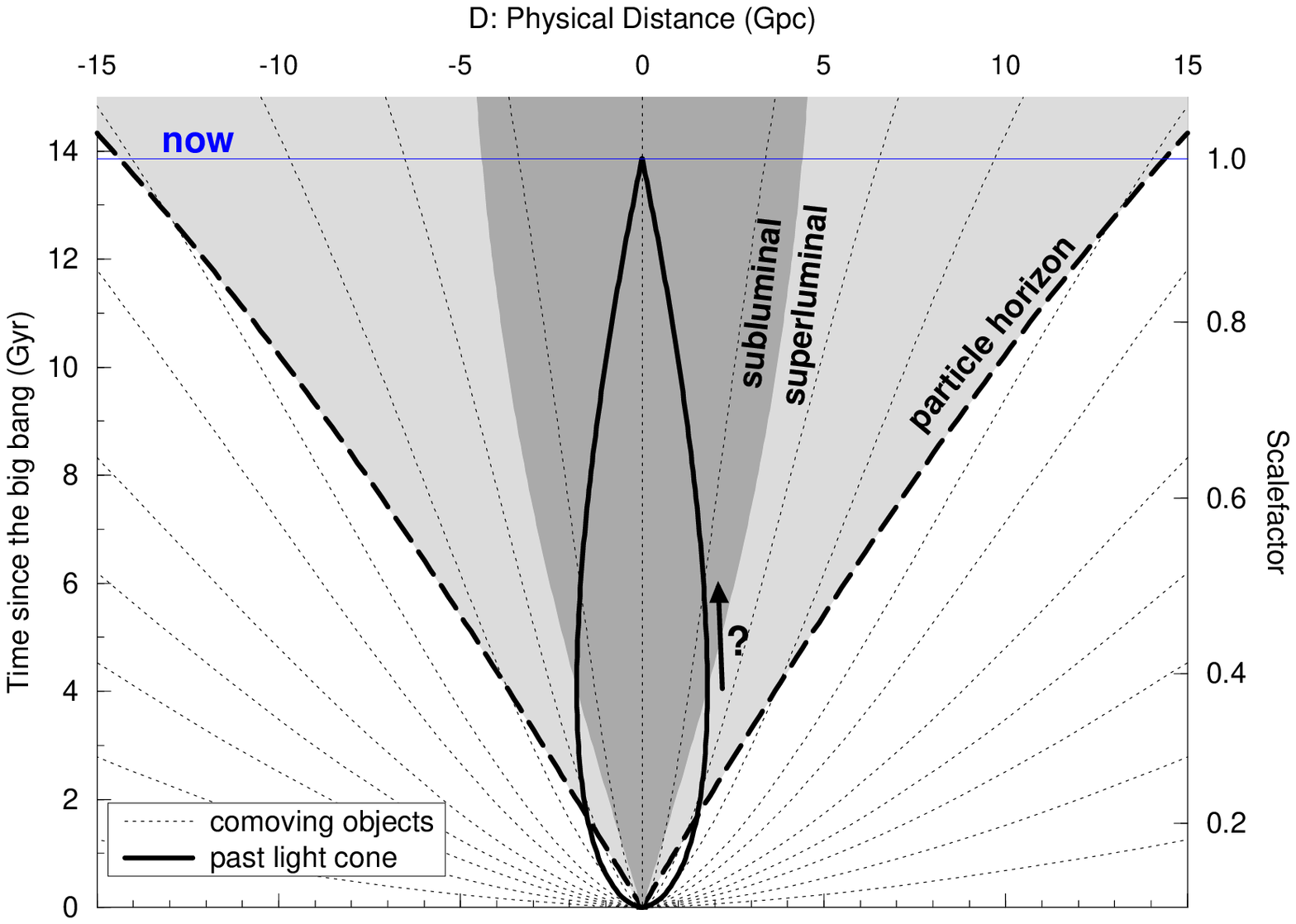,height=3.45in,width=6in}}\vspace{-0.8in}
\centerline{\epsfig{file=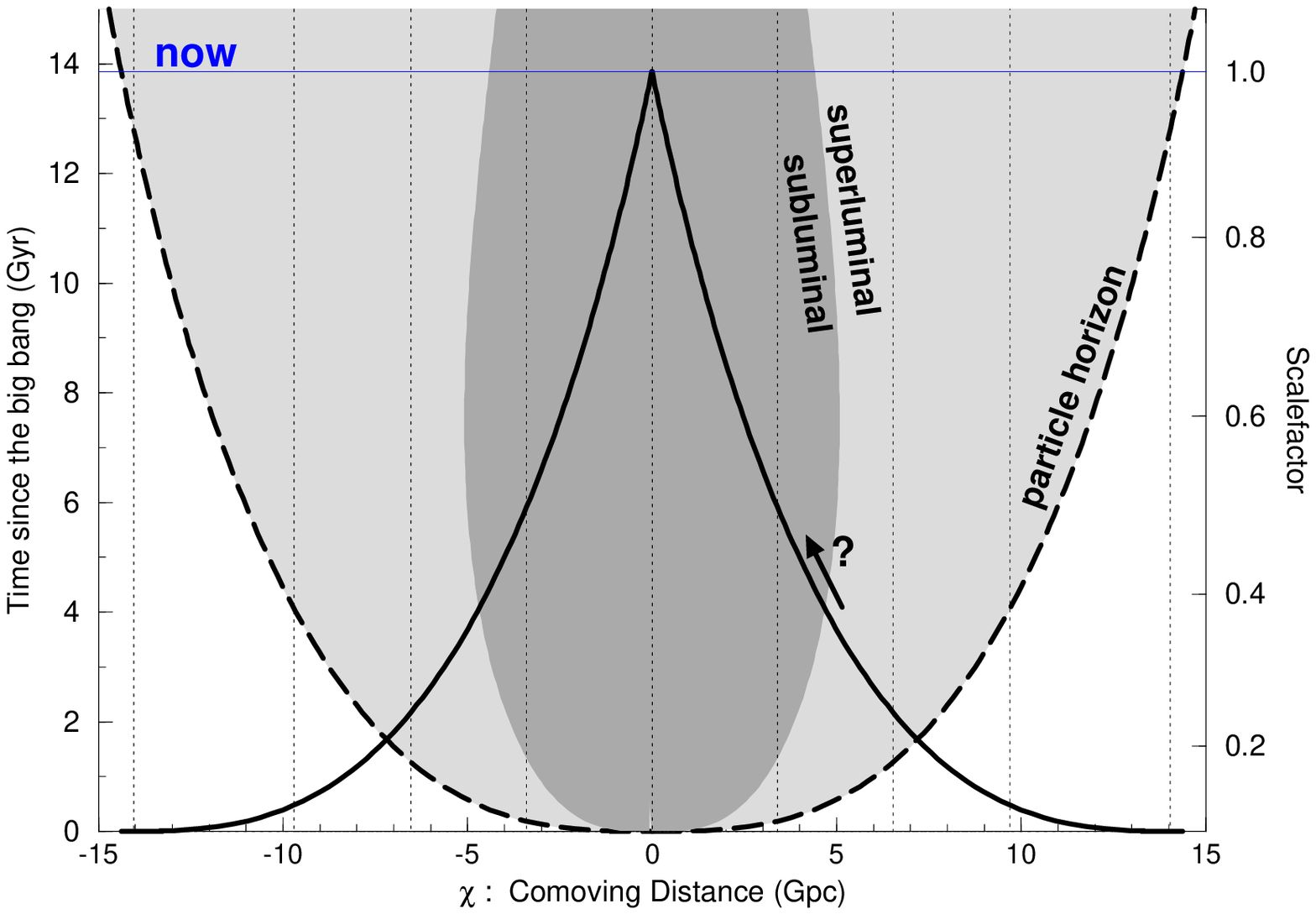,height=3.45in,width=6in}}\vspace{-20pt}
\caption{These spacetime diagrams are generated by inserting the parameters $(h,\Omega_M,\Omega_{\Lambda})=(0.68,0.3,0.7)$ into 
Friedmann's equation.  The upper panel is in physical distance and has a tear-drop-shaped past lightcone. The lower panel is 
in comoving distance and has a trumpet-shaped past light cone.   The past lightcones enclose the events we have been able to see.
Constant time surfaces are horizontal slices.  
On a surface of constant time the dark grey shading encloses the region of space which is receding subluminally; 
all other regions are receding superluminally. The light grey shading defines the observable universe and is bounded by
the particle horizon. 
Following the worldline of a comoving galaxy that was receding superluminally at the time of emission, shows that the galaxy is {\em still} receding superluminally.  In the viable GR models of Fig. [1], all objects with a redshift greater than $\sim 3$ are, {\em and always have been}, receding faster than the speed of light. 
How can photons reach us from regions of space that are receding superluminally?
How can they cross from the light grey into the dark grey? 
(see text).
}
\label{F:tdavis:fig2}
\end{figure}

In Figure~\ref{F:tdavis:fig2} the boundary between the light and dark grey regions is the Hubble sphere, the distance, $D_{HS}(t)=c/H(t)$, at which galaxies are receding at the speed of light.
The comoving distance to the Hubble sphere increases when the universe decelerates and decreases when the universe accelerates. The Hubble sphere is not an horizon of any kind; it passes over particles and photons in both directions.

How can photons move from the light grey superluminal region into the dark grey subluminal region? How can a swimmer make headway against a current that is faster than she can swim?
It seems impossible. 
First consider the problem in comoving coordinates. 
Photons propagate towards us along our trumpet-shaped past lightcone with a comoving velocity $\dot{\chi}_{\gamma}=-c/a$. Therefore, photons move inexorably through comoving space, irrespective of their position in relation to our Hubble sphere. Their comoving velocity is always negative. Their comoving distance always decreases.

Now consider the problem in physical space.
The tear-drop shape of our past lightcone means that the distance between us and the most distant photons we now see, was once increasing.
The relevant quantity for understanding this behaviour is the total velocity of a photon that is heading
towards us:  $v_{tot} = v_{rec} -c = HD -c = \dot{a} \chi_{\gamma} -c$. 
The total velocity of distant photons is not constant
because it is the sum of the distance-dependent recession velocity ($v_{rec}$) 
and the constant peculiar velocity, $c$.  
When $\dot{a} \chi_{\gamma} > c$ the distance between us and the photon increases.

Photons now reaching us from our particle horizon were emitted at the Big Bang. Since $\lim_{t\rightarrow0}\dot{a}=\infty$ in all viable cosmological models, $\dot{a} \chi_{\gamma} > c$ would have been satisfied and these photons would have inially receded from us.  Similarly, the first photon we receive from {\em any} object was emitted from a region with $\dot{a} \chi_{\gamma}>c$, $v_{tot}>0$.  In the early universe as time progresses, both $\chi_{\gamma}$ and $\dot{a}$ decrease.  Thus, $v_{tot}$ of these photons evolves from positive to negative, and the teardrop shape of our physical past lightcone is ubiquitous to all times.   

%Consider an initial $\chi_{\gamma}$ for which $\dot{a} \chi_{\gamma} > c$ is true. Since $\lim_{t\rightarrow0}\dot{a}=\infty$ in all viable cosmological models an arbitrarily small $\chi_{\gamma}$ satisfies this condition, i.e., all regions were once receding faster than the speed of light, i.e., the dark grey subluminal region gets smaller in comoving coordinates as $t\rightarrow0$.  In the early universe as time progresses, both $\chi_{\gamma}$ and $\dot{a}$ decrease.  Thus, $v_{tot}$ will evolve from positive to negative resulting in the teardrop shape of our physical past lightcone.  

%To make the swimming analogy work we need to have a swimmer that maintains a constant speed, mark comoving coordinates on the water and make the speed of the current ($v_{tot}$) depend on where in the water the swimmer is.  
%As she moves forwards with respect to the water (makes headway in comoving coordinates) she passes the point at which the water's speed matches her own, after which she can make headway with respect to the banks.  Unfortunately for her she has been swept backwards, and now has a much further distance to swim.  This is analogous to the photon whose distance initially increases, but which finally approaches us after it has moved through comoving coordinates to a region of space that is receding subluminally.  

The swimming analogy fails because, unlike recession velocity which is smaller
 at smaller comoving distances, the current the swimmer has to face is the {\em
 same} at all comoving distances.  Our swimmer has to battle an unrelenting cur
rent, while the photon constantly moves into regions with a slower ``current'' 
(slower $v_{rec}$).

\end{document}